\documentclass[aip,graphicx]{revtex4-1}
\usepackage{graphicx}
\usepackage{dcolumn}% Align table columns on decimal point
\usepackage{bm}% bold math
\usepackage[utf8]{inputenc}
\usepackage[T1]{fontenc}
\usepackage{mathptmx}
\usepackage{etoolbox}
\usepackage{color}
\draft % marks overfull lines with a black rule on the right

\begin{document}

%\preprint{}

\title{Experimental Realization of Scanning Quantum Microscopy} %Title of paper



%\author{}
%\thanks{}
%\affiliation{}

\author{V. F. Gili}
\email{valerio.gili@uni-jena.de}
\affiliation{Friedrich-Schiller-Universität Jena, Institute of Applied Physics, Abbe Center of Photonics, 07745 Jena, Germany}
\author{C. Piccinini}

\affiliation{Friedrich-Schiller-Universität Jena, Institute of Applied Physics, Abbe Center of Photonics, 07745 Jena, Germany}
\affiliation{Politecnico di Milano, Dipartimento di Fisica, 20133 Milano, Italy.}

\author{M. Safari Arabi}
\affiliation{Friedrich-Schiller-Universität Jena, Institute of Applied Physics, Abbe Center of Photonics, 07745 Jena, Germany}
\author{P. Kumar}
\affiliation{Friedrich-Schiller-Universität Jena, Institute of Applied Physics, Abbe Center of Photonics, 07745 Jena, Germany}
\author{V. Besaga}
\affiliation{Friedrich-Schiller-Universität Jena, Institute of Applied Physics, Abbe Center of Photonics, 07745 Jena, Germany}
\author{E. Brambila}
\affiliation{Friedrich-Schiller-Universität Jena, Institute of Applied Physics, Abbe Center of Photonics, 07745 Jena, Germany}
\affiliation{Fraunhofer Institute for Applied Optics and Precision Engineering, 07745 Jena, Germany.}
\author{M. Gräfe}
\affiliation{Friedrich-Schiller-Universität Jena, Institute of Applied Physics, Abbe Center of Photonics, 07745 Jena, Germany}
\affiliation{Fraunhofer Institute for Applied Optics and Precision Engineering, 07745 Jena, Germany.}
\author{T. Pertsch}
\affiliation{Friedrich-Schiller-Universität Jena, Institute of Applied Physics, Abbe Center of Photonics, 07745 Jena, Germany}
\affiliation{Fraunhofer Institute for Applied Optics and Precision Engineering, 07745 Jena, Germany.}
\author{F. Setzpfandt}
\affiliation{Friedrich-Schiller-Universität Jena, Institute of Applied Physics, Abbe Center of Photonics, 07745 Jena, Germany}
\affiliation{Fraunhofer Institute for Applied Optics and Precision Engineering, 07745 Jena, Germany.}


\begin{abstract}
Quantum imaging is an ever expanding research field, in which the aim is to exploit the quantum nature of light to enhance image reconstruction capabilities. %This approach has led to the demonstration of remarkable results, such as sub-shot noise imaging, ranging, two-colour imaging, quantum-enhanced signal-to-noise ratio (SNR) and others. 
Despite a number of successful demonstrations for quantum imaging, quantum microscopy still seems out of the range for practical applications, due to different physical and technical reasons.
%since in wide-field approaches the number of independent spatial modes allowed by spontaneous parametric down-conversion (SPDC) photon-pairs fundamentally limits the achievable resolution, while scanning approaches have been reported only for sample raster-scanning, typically suffering from slow speeds. 
Here we propose an imaging method exploiting the quantum correlations of photon pairs and a scanning microscope to achieve fast, single mode quantum imaging. We first test our technique on a metal grating to estimate the resolution capabilities of our system. Moreover, we asses its potential in terms of the number of available independent pixels at full resolution compared to different quantum imaging approaches. Finally, we demonstrate scanning quantum microscopy of onion epithelial cells, paving the way towards scalable quantum microscopy for bio-physical applications. Our results, combined with the rapidly evolving photon-pair generation and detection technology towards the mid-infrared, could lead to an extension of quantum microscopy applications towards the mid-infrared, to access the molecular fingerprint region. 
\end{abstract}


\maketitle %\maketitle must follow title, authors, abstract and \pacs


Quantum correlations, ever since the formulation of the EPR paradox \cite{epr}, are at the core of the observation of many counter-intuitive effects, which have constantly stimulated research towards their exploitation in new applicable technologies. In particular, quantum correlations of light have been predicted theoretically to unlock quantum advantages in imaging and sensing applications with reduced noise and increased contrast \cite{theory, genovese}. Thus, they have been employed for a number of imaging methods, such as quantum illumination \cite{QI, expQI}, quantum ghost imaging \cite{klyshko, QGI}, imaging with undetected photons \cite{Lemos}, as well as spectroscopy and sensing \cite{Kriv, spectroscopy}. Following first proof-of-principle demonstrations, remarkable results have been achieved in applying such quantum protocols to microscopy, ranging from improved signal-to-noise ratio (SNR) in coherent Raman microscopy \cite{QRaman}, in quantum ghost imaging  of low-transmission samples \cite{DGI}, in differential interference-contrast microscopy \cite{ent_mic}, and in quantum illumination imaging \cite{QII}; sub-shot noise quantum imaging \cite{ssnim1, ssnim2, ssnim3, ssnim4} and absorption spectroscopy \cite{ssnsp1, ssnsp2}; quantum illumination-based noise radar \cite{QInr} and ranging \cite{Qlidar}. Despite the numerous promising results obtained so far, the scaling of the demonstrated systems towards practical applications is still challenging. This is on one hand due to the limited amount of independent spatial modes generated by typically employed photon-pair sources based on spontaneous parametric down conversion (SPDC), which fundamentally limits the amount of spatial information that can be contained in a wide-field quantum image \cite{ssnim4}. In typical wide-field quantum imaging schemes, the spatially resolving detector does not detect the photons interacting with the object and the spatial information is retrieved using the spatial correlations between the two photons of a pair. On the other hand, for imaging techniques based on correlation measurements, and in particular for quantum ghost imaging, a number of severe technical limitations apply. To obtain correlations, single-photon counting and time-stamping are needed. In wide-field imaging, this necessitates the use of intensified cameras \cite{isnp}. However, these systems need to temporally delay the optical quantum states while maintaining the image, which has been achieved with $\sim{30}$ m-long image-preserving delay lines \cite{isnp} that severely hinder practical application, or with complex quantum memory systems \cite{varsavia} that require great efforts for atom cooling and trapping. Very recently, the use of different image sensors with single-photon timing capability for quantum imaging applications was demonstrated \cite{FBK, faccio, timepix, SPADGI}, which overcome the above-mentioned limitations of ICCD cameras. However, correlations were only demonstrated between the pixels of one sensor, limiting the applicability of these devices. Alternatively, scanning imaging is possible, but has been implemented only by raster-scanning the sample, thus limiting the acquisition speed \cite{ssnim1, ssnim2, SPADGI}.
In this work, we propose and implement a method based on fast illumination scanning with galvanic mirrors, to overcome limitations on the amount of obtainable spatial information and acquisition speed of previously demonstrated correlation-based quantum imaging schemes. Scanning imaging provides the fundamental advantage of obtaining spatial information in a sequential manner, hence allowing for a spatially single mode approach. In this configuration, we can employ high-efficiency and single-mode integrated SPDC sources and single-photon detectors, and we can benefit from the fast scanning technology implemented in modern microscopes. Thus, compared to earlier approaches for correlation-based quantum imaging, we do not have to resort to image-preserving delay lines or slow scanning techniques involving step-wise movement of the sample, with the ultimate goal of leading the way towards scalable quantum-enhanced microscopy for bio-imaging applications.
\begin{figure}
\includegraphics[width=\textwidth,keepaspectratio]{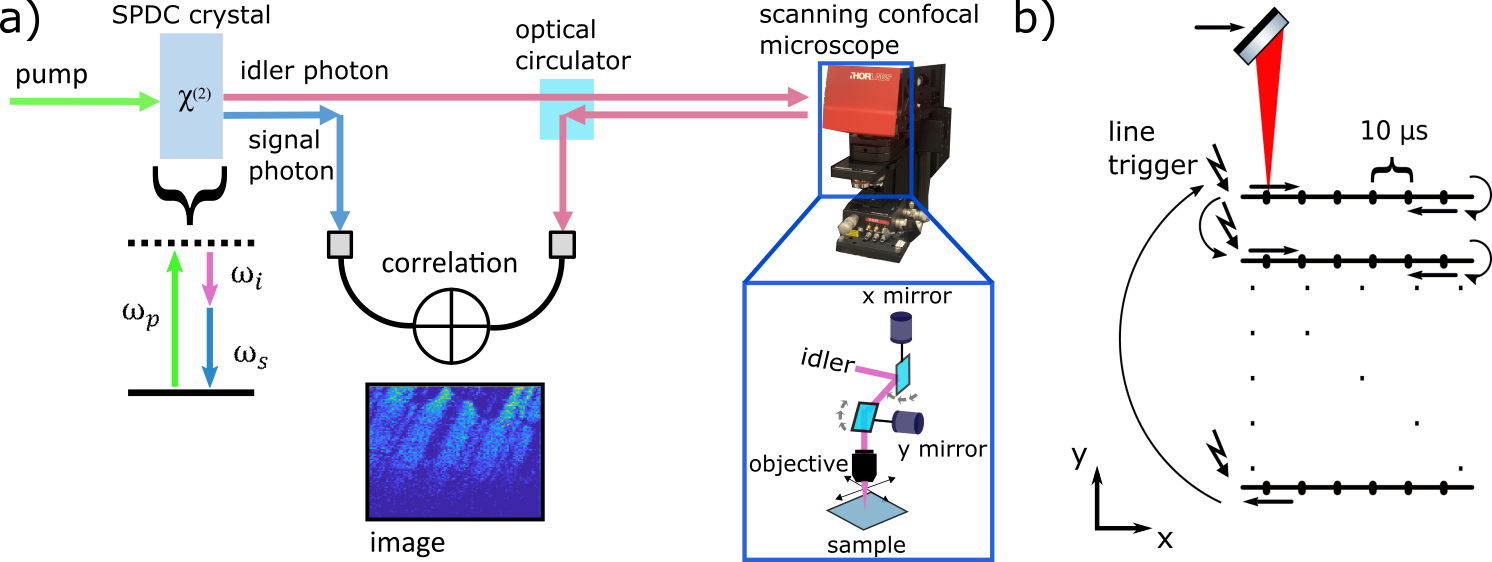}
\caption{\label{fig:SPDC} a) Schematics of our setup for scanning quantum microscopy: a pump beam impinges on a nonlinear crystal to produce non-degenerate SPDC. After splitting, the shorter photon is directly sent to a single-pixel detector, while the other is coupled into a scanning microscope. Reflected photons are collected through the same path, separated from input ones with a circulator, and detected by another single-pixel detector to reconstruct quantum correlations. b) Sketch of the  scanning procedure used to reconstruct quantum correlations: at the beginning of each line the scanning head sends a line trigger to our time-tagger to assign coincidence counts to every pixel, then the x-mirror scans the sample at constant speed. Once the trigger reaches the line end, it decelerates, turns back, and scans the same line in the opposite direction. Once the second scan of the first line is complete, the y-mirror moves to the next line and a new line trigger is produced. Once the frame is complete, the galvo mirrors move back to the initial position during a time interval set to be equal to one frame scan.}
\end{figure}
Our approach to scanning quantum imaging is conceptually shown in Fig.~\ref{fig:SPDC}a. Correlated photon pairs are generated through non-degenerate SPDC in a second-order nonlinear crystal, and subsequently split with a dichroic beam-splitter. The signal photons with shorter wavelength are directly sent to detection. The long-wavelength idler photons are coupled to a commercially available scanning microscope (Thorlabs CM100) and scanned across the sample surface with a pair of galvo mirrors for the x- and y-directions. Photons reflected from the sample are collected through the same objective and propagate along the same path, until they get separated from the input photons with a Faraday rotator and a polarization beam-splitter, which sends them to detection. Finally, correlations between signal and idler detection events allow to reconstruct the quantum image. 
Importantly, as the image is obtained point-by-point, it is sufficient to generate both signal and idler photons in a single optical mode. 
%Multi-mode spatial correlations between signal and idler are not necessary. 
This allows the use of nonlinear waveguides as photon-pair source, which enable enhanced generation efficiencies compared to bulk nonlinear crystals due to their small mode size and long propagation length \cite{FOP}. Furthermore, light generated in single-mode waveguides can be sent to fiber-coupled single-photon detectors with high efficiency. In our implementation, the SPDC source is a 50~mm-long Ti-indiffused periodically poled lithium niobate (PPLN) waveguide designed for Type-0 phase matching, able to produce non-degenerate photon pairs in the spectral range 1.1 $\mu$m - 2.5 $\mu$m \cite{frank, pawan}. %The choice of single-mode waveguide-based SPDC is motivated by the fact that smaller mode volumes combined with a longer non-linear crystal sample, compared to bulk crystals, allow for greater nonlinear generation efficiencies \cite{FOP}, while at the same time no more than one independent k-vector is needed in the scanning regime, since the maximum allowed pixel number is set by the SM scanning capabilities and the resolution is limited by the confocal nature of this microscope. Furthermore, SM do not require image-preserving delay lines and are generally faster than their raster-scanning analogues \cite{ssnim1}. 
We produce non-degenerate photon pairs by pumping our waveguide at 772.3 nm, and generate signal photons at 1435 nm, measured through an InGaAs spectrometer. This corresponds to an idler wavelength of 1673 nm, calculated with the wavelength correlation that stems from SPDC energy conservation:  
\begin{equation}
    \lambda_i^{-1}=\lambda_p^{-1}-\lambda_s^{-1},
\end{equation}

where $\lambda_{p,s,i}$ is the wavelength of pump, signal, and idler photons, respectively. Concerning photon detection, since our imaging approach is single mode, we chose superconducting nanowire single-photon detectors (SNSPDs - Single Quantum EOS), in which photons to be detected are coupled through single-mode fibers. Compared to other quantum imaging implementations, in which photons are typically coupled to multi-mode fibers, the lower coupling efficiency to single-mode fibers is outweighed by the high detection efficiency of our SNSPDs over a broad spectral range in the near-infrared: 0.85 at 1.31 $\mu$m, 0.81 at 1.55 $\mu$m, and 0.6 at 1.86 $\mu$m. This detection range is broader than for the InGaAs detectors used in earlier implementation and our experiments conducted at 1673~nm are at the edge of the detection bandwidth of cooled InGaAs systems. Furthermore, SNSPDs for larger wavelengths are under development \cite{snspd}, which could enhance the usefulness of our approach for infrared quantum imaging. Notably, in our approach, the illumination scanning mechanism alleviates the requirements on the source correlations. In such way, the limited resolution stemming from the finite correlation width of SPDC sources becomes irrelevant for resolution purposes \cite{elkin}, and imaging with single-mode sources becomes possible.

\begin{figure*}[htb]
\centering
\includegraphics[width=\textwidth,keepaspectratio]{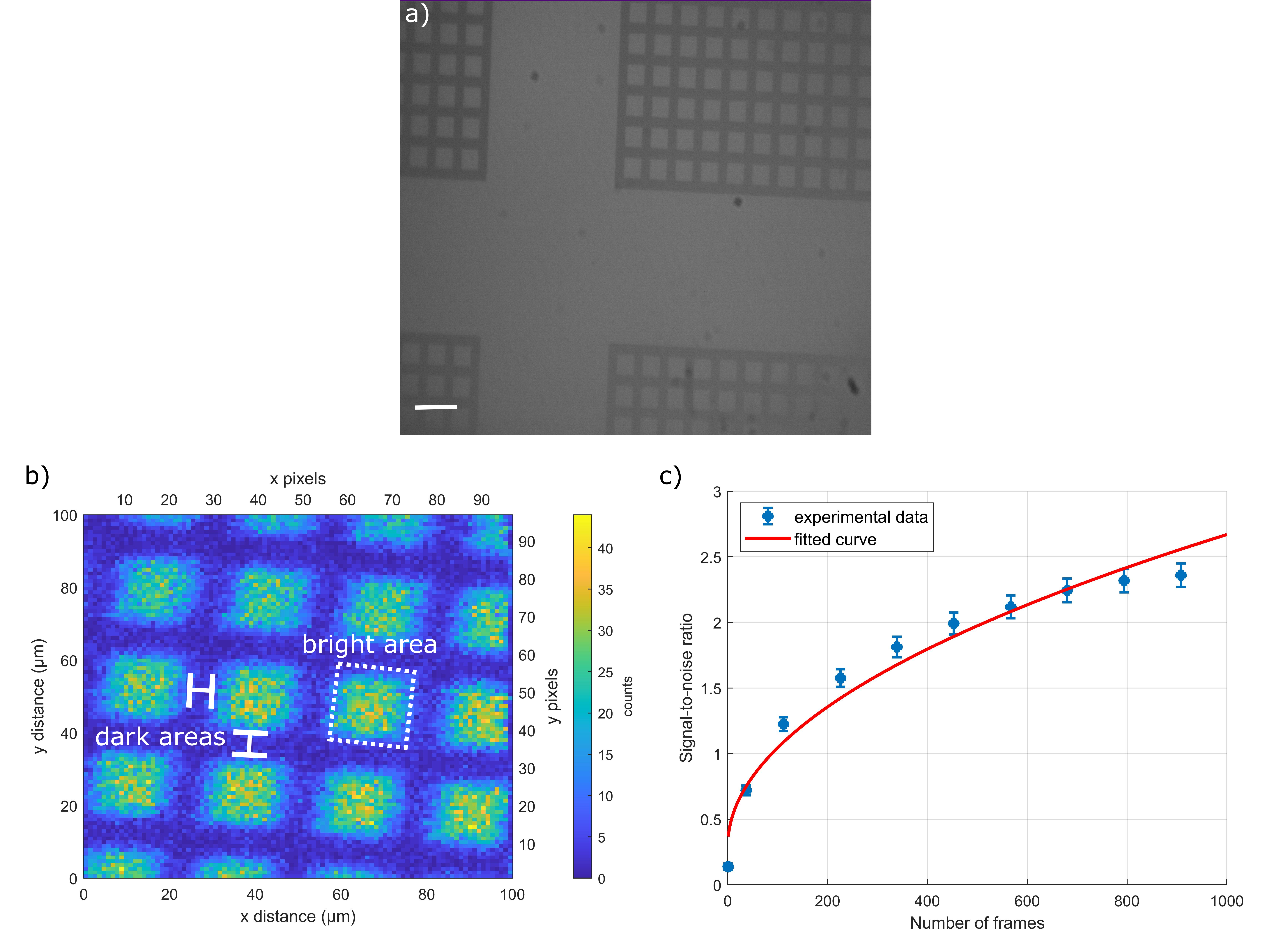}
\caption{\label{fig:setup} a) Wide-field white-light image of the silver grating sample investigated. The white bar is 50 $\mu$m wide, while the squares' size is 20 $\mu$m and their separation is 10 $\mu$m. b) Quantum image of the same sample obtained with a 0.3 NA reflective objective in a 4 min integration time. The scanned area is 100 $\mu$m x 100 $\mu$m. The colorbar on the right side relates to the number of coincidences detected per pixel. c) Experimental SNR (blue dots) calculated with Eq. 2 vs the number of accumulated frames. Importantly, the left-most experimental point corresponds to a frame number of 1. The result is fitted with a function of the form $f(x)=A\cdot \sqrt x$ (red solid line), with a $R^{2}=0.93$.}
\end{figure*}

Image information in our microscope is obtained by scanning the stream of idler photons across the sample. The scanning procedure is performed according to the scheme in Fig.~\ref{fig:SPDC}b. Once the scanning starts, the scanning mirror accelerates to move the focal point across the sample until it reaches a constant speed, which is user-defined once total pixel number, scan area, and pixel-dwell-time (10 $\mu$s) are specified in the microscope software. Once a plateau scan speed is reached, the scanner produces a first line trigger, and the x-mirror starts scanning the first line. %, with a pixel dwell time set to 10 $\mu$s, and a total pixel number set by the user. 
Once the end of the line is reached, the scanner takes a fixed amount of time of 400 $\mu$s to decelerate to speed zero, turn around, and re-accelerate to the plateau constant speed. This is referred to as "turnaround time", and photons detected during these intervals must be discarded since the motion law of the scanner is not known. After one turnaround time, the same line gets scanned again in the opposite direction. Once the second scan of the first line is over, the scanner turns around again, moves one step in the y-direction, and once the plateau scan speed is reached a second line trigger is sent. This procedure repeats until the last line gets scanned, defining the end of one frame. Once every frame is complete, the scanner takes a fixed time interval equal to the frame scan duration to return to the initial position, called "flyback time". Analogously to the turnaround time, photon counting events occurring during the flyback time have to be discarded. The described scan procedure is completely independent from the excitation, which creates photon pairs randomly distributed in time. However, the timing information given by the line triggers together with the knowledge on turnaround and flyback times enables to reconstruct images from the recorded continuous stream of correlations.
To this end, a Time-to-Digital Converter (IDQ800, ID quantique) records timetags corresponding to signal and idler photon detection events together with the line trigger signals from the scanner. First, the photon-pair correlation histogram is reconstructed, so as to measure the time delay induced by the optical path difference between signal and idler photons, shift the single photon timetags accordingly, and generate an array of coincidence timetags. These can then be assigned to spatial pixels by discretizing the stream of coincidences into time-bins of 10~$\mu$s length. The correlations within each time-bin are then assigned to a single image pixel. The temporal length of the pixels is chosen to match the distance the focus travels in 10~$\mu$s, the pixel-dwell-time.

In order to demonstrate the imaging capabilities of our system, we captured the quantum coincidence image of a silver grating, consisting of periodic square structures resting on a glass substrate. We first installed a 0.3 NA (15X) reflective objective, chosen to minimize losses, and aligned the sample with a white-light source integrated into our scanning microscope (Fig.~\ref{fig:setup}a). We then selected a sample zone which includes squares of nominal size of 20 $\mu$m x 20 $\mu$m and 10 $\mu$m separation, and set the scanning area to 100 $\mu$m x 100 $\mu$m. The resulting image after a 4 min integration time, containing data from 908 individual scan frames, is shown in Fig.~\ref{fig:setup}b. %: x- and y- axes represent x- and y- pixels respectively, while the z- axis of the colormap represents the coincidence counts.
Each image pixel corresponds to $\sim$1.04 $\mu$m x 1.04 $\mu$m, determined from the scan speed and confirmed by referencing with the known nominal sizes of the measured silver squares. The capability to reconstruct an image from this large number of frames demonstrates that our reconstruction algorithm correctly incorporates the different time constants of the scanner system, as otherwise each frame would be shifted with respect to the others and the overall image would be blurred.

\begin{figure*}[htb]
\centering
\includegraphics[width=\textwidth,keepaspectratio]{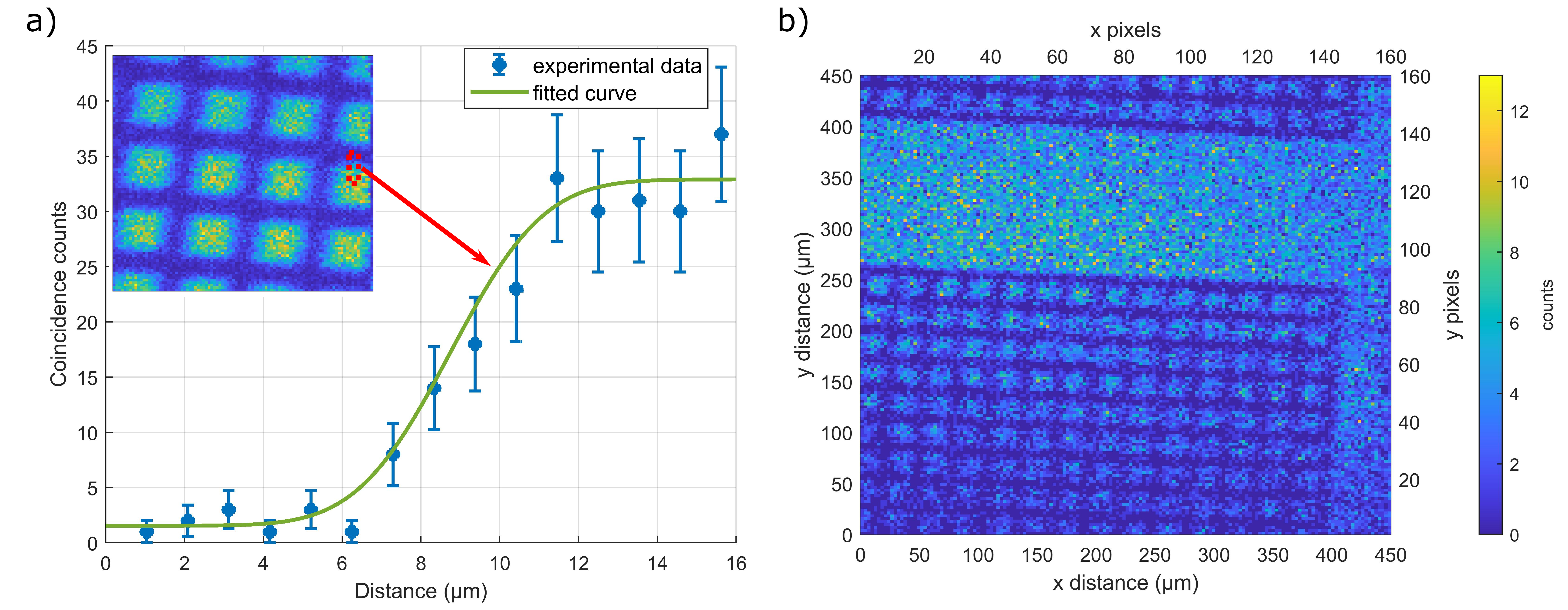}
\caption{\label{fig:resolution} a) Explanatory error function fit of a silver square edge extracted from Fig. 2a (see inset) to estimate the system resolution. Experimental points are indicated with blue dots, with error bars calculated assuming Poissonian detection statistics, while the green solid line represents the error function fit (Eq. 3). b) Quantum image at full resolution conditions: the scanned area is 450 $\mu$m x 450 $\mu$m in size, and the pixel number is 160 x 160. The bright area represents the separation zone between identical gratings. The colorbar indicates the number of coincidence events per pixel used to reconstruct the image.}
\end{figure*}

To assess, how the image quality depends on the number of frames, we evaluated the SNR defined as\cite{genovese}: 
\begin{equation}
    \mathrm{SNR}=\frac{|\langle C_{bright}-C_{dark} \rangle|}{\sqrt{\langle \delta^{2}(C_{bright}-C_{dark}) \rangle}},
\end{equation}
where $\delta C$ is the coincidence count fluctuation and $C_{bright/dark}$ are the coincidence counts of the reconstructed image in the bright or the dark areas of the sample, respectively. We defined these areas based on the average counts per pixel recorded in the image reconstructed from idler photons, which we assume to be our ground truth, where bright areas have a count number above, dark areas a count number below the average. Thus, the bright areas correspond to the silver squares and the dark areas to the gaps between them, as schematically shown in Fig.~\ref{fig:setup}b. We then calculated the SNR for different frame numbers.  
The result is reported in Fig.~\ref{fig:setup}c (blue dots). Error bars are included assuming Poissonian detection statistics. As we theoretically expect the SNR to grow with the square root of the frame number \cite{genovese2}, we fit the data with a function of the form $f(x)=A\cdot \sqrt{x}$ (solid red line), where $A$ is a free parameter. Overall, our experimental results shows a good agreement with the square-root dependence, with $R^{2}=0.93$, where slight deviations may be due to the fact that the photon number is not exactly the same in every frame.

We next tested the spatial resolution of our scanning quantum microscope and compared it with the confocal resolution limit. To do this, we took the quantum image obtained with the 0.3 NA objective (Fig. 2a) and sampled 20 linescans in the transition zone between glass (dark area) and silver (bright area), so as to effectively investigate the edge response of our scanning microscope \cite{confocal}. The resolution then was estimated by fitting the line profiles with an error function convoluted with a Gaussian curve as\cite{resolution}   
\begin{equation}
    f(x)=A\cdot \left[ erf\left( \frac{x-(\mu-a)}{\sqrt{2}\sigma_{res}} \right) \right] +B,
\end{equation}
where $A$, $B$ are free parameters, $x$ is the spatial coordinate, $\mu$ is the mean value, $\sigma_{res}$ is the variance that constitutes the resolution measure we adopted. Fig.~\ref{fig:resolution}a shows an explanatory error function fit of one of the edge line profiles extracted from the full quantum image of Fig.~\ref{fig:setup}b. Repeating the fit procedure for all sampled linescans, we obtained an average resolution of $ 2.00 \pm 0.29$ $\mu$m. This value is consistent with the 20$\%$ - 80$\%$ edge-response criterion for confocal microscopes, estimated as 0.33$\cdot\lambda /$NA$=1.87$ $\mu$m \cite{confocal}.  We note, that the resolution criterion we adopted is not independent from single-point (point-spread function) and two-point (Rayleigh) resolution criteria since the system point-spread function is the first derivative of the edge-response function \cite{resolution2}. A similar resolution analysis was repeated for another reflective objective with 0.5 NA (40X) and we obtained an average resolution of $1.25 \pm 0.16$ $\mu$m, yet again consistent with the confocal edge-response limit, which amounts to 1.12 $\mu$m in this case.   
Importantly, the resolution of our microscope is only limited by the employed optics, contrary to wide-field quantum imaging methods the spatial properties of the used photon-pair source are not limiting. This also lifts the restriction on the number of individually resolvable elements in one image, that inevitably connects the achievable field of view with the spatial resolution in wide-field quantum imaging \cite{ssnim4}. To demonstrate this, we increased the field of view scanned by the microscope to 450 $\mu$m x 450 $\mu$m. Using the 15X objective as before, we imaged portions of two of the silver gratings. The obtained image is shown in Fig.\ref{fig:resolution}b. The number of image pixels is 160 x 160 with a spatial pixel size of about 2.8 $\mu$m x 2.8 $\mu$m, slightly larger than the resolution limit due technical constraints for choosing the parameters of the scanner and reconstruction algorithm. Nevertheless, the image contains 25600 independent elements and constitutes a proof-of-principle demonstration of the capabilities of our method, compared to wide-field quantum microscopy, in which a maximum of $\sim$ 8000 independent pixels at a resolution of 5 $\mu$m could so far be exploited \cite{ssnim4}, and undetected-photon microscopy, demonstrated with up to 900 independent spatial modes \cite{ramelow}.

\begin{figure*}[htb]
\centering
\includegraphics[width=\textwidth,keepaspectratio]{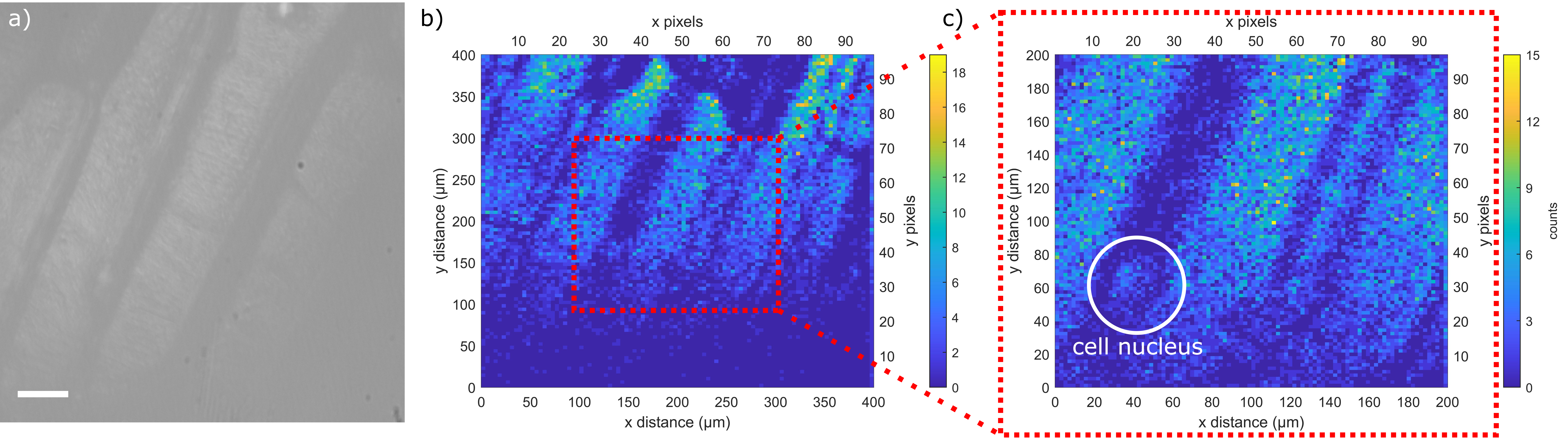}
\caption{\label{fig:onion} a) Widefield white light image of the onion cell sample under investigation. The white bar corresponds to a 50 $\mu$m length. b) Quantum image of onion epidermal cells with 400 $\mu$m x 400 $\mu$m size accumulated over 4 min. c) 2X-zoomed scan of the same area, allowing to identify the detail of a cell nucleus, obtained over the same integration time. }
\end{figure*}

To demonstrate the potential of our quantum imaging approach for bio-imaging applications in the infrared, we finally imaged a sample consisting of onion epidermal cells. The sample was prepared in the following way \cite{onion}: we cut an onion and peeled off a piece of the membrane that separates one layer from the other. We then placed the sample over a silver mirror, added a drop of iodine solution, and finally covered the sample with a microscope slide. 
The choice of this sample is motivated by the fact that these kind of cells are relatively big, with an average transverse width of 50 $\mu$m - 75 $\mu$m \cite{onion}, and hence constitute an ideal benchmark sample for our scopes. We first acquired a 400 $\mu$m x 400 $\mu$m wide quantum image in 4 min accumulation time (Fig. 4a), in which onion epidermal cells can be clearly distinguished, and then performed a 2X-zoom of the same area (200 $\mu$m x 200 $\mu$m) that allowed us to resolve a cell nucleus (Fig. 4b). From the 2X image we can infer a nucleus dimension of $\sim20$ $\mu$m, which is in line with typical observations \cite{onion-nucleus}.
With this work, we experimentally demonstrate scanning quantum microscopy. 
%Based on non-degenerated SPDC in a PPLN waveguide photon pairs are split according to their wavelength. While one acts as a herald, its partner is coupled into galvo-galvo scanning microscope illuminating a sample. 
After characterization of SNR and resolution capabilities, our scanning quantum microscope is employed to epithelial tissue of an onion demonstrating its feasibility for bio-imaging. Our approach benefits from several points: i) as single-mode confocal technique we do not suffer from the limited number of modes that can be generated via SPDC and ii) principally can fully exploit the confocality for three-dimensional quantum imaging. iii) The inherent low photon flux can become of importance in bio-imaging~\cite{marta}, where high fluxes could induce photo-damage and cause unwanted heating of the surrounding media, and in eye-security applications~\cite{isnp}. iv) There is no need of an image preserving delay line nor computational challenging post selection of coincidence events in an asynchronous detection scheme. With these advantages we pave the way towards scalable plug-and-play quantum microscopy, with the potential to reach a sub-shot-noise microscopy regime. Currently, our proof-of-principle measurement device does not yet provide a quantum advantage, due to the small coupling efficiencies through the entire microscope system. However, by improving all optical elements, e.g. by using an SPDC source with modes matched to the input fibers of the SNSPD and optimizing the optical design of the microscope, the losses can be significantly decreased and reaching a quantum advantage seems feasible.
As direct outlook, hyperspectral imaging in the NIR spectral portion of 1.2 $\mu$m - 2.3 $\mu$m is potentially accessible by our SPDC source~\cite{pawan}. In this spectral range a number of vibrational modes, associated to meaningful compounds like proteins and carbohydrates, could be investigated~\cite{bio-app}. Furthermore, sources of photon pairs with signal photons at much longer wavelengths have been demonstrated~\cite{mohit} and suitable single-photon detectors for such low energies are under development~\cite{snspd}. Hence, an extension of the spectrum for scanning quantum imaging towards the mid-infrared to access the molecular fingerprint region seems feasible. Conversely, our technique can also be extended to the visible, where photon detection technology is more developed, and interesting bio-applications of quantum imaging can be identified, such as DNA concentration measurements \cite{DNA}.

% If in two-column mode, this environment will change to single-column format so that long equations can be displayed. 
%\begin{widetext}
%$$\mbox{put long equation here}$$
%\end{widetext}

% Figures should be put into the text as floats. 
% Use the graphics or graphicx packages (distributed with LaTeX2e).
% Put the label that you will use with \ref{} command in the braces of the \label{} command.
%
% \begin{figure}
% \end{figure}

% \begin{table}
% \end{tabular}
% \end{table}

%\begin{acknowledgments}
% Put your acknowledgments here.
%\end{acknowledgments}
\begin{acknowledgments}
V. F. G. acknowledges M. Kumar, for fruitful discussions, M. Gilaberte-Basset for help in sample preparation. This work was supported by the Thuringian Ministry for Economy, Science, and Digital Society and the European Social Funds (2021 FGI 0043); European Union’s Horizon 2020 research and innovation programme (Grant Agreement No. 899580); the German Federal Ministry of Education and Research (FKZ 13N14877, 13N15956); and the Cluster of Excellence ‘‘Balance of the Microverse’’ (EXC 2051 – project 390713860). 
This article may be downloaded for personal use only. Any other use requires prior permission of the author and AIP Publishing. This article appeared in "V. F. Gili, C. Piccinini, M. Safari Arabi, P. Kumar, V. Besaga, E. Brambila, M. Gräfe, T. Pertsch, F. Setzpfandt; Experimental realization of scanning quantum microscopy. Appl. Phys. Lett. 5 September 2022; 121 (10): 104002. https://doi.org/10.1063/5.0095972" and may be found at "https://pubs.aip.org/aip/apl/article-abstract/121/10/104002/2834301/Experimental-realization-of-scanning-quantum?redirectedFrom=fulltext". Copyright 2022 Author(s). This article is distributed under a Creative Commons Attribution-NonCommercial 4.0 International (CC BY-NC) License. https://creativecommons.org/licenses/by-nc/4.0/.
\end{acknowledgments}

\section*{Data Availability Statement}

The data that support the findings of this study are available from the corresponding author upon reasonable request.

\nocite{*}

\bibliography{biblio}

\end{document}